\documentclass[aps,prc,floats,floatfix,preprint,superscriptaddress,nofootinbib]{revtex4}  
\def\be{\begin{eqnarray}}    
\def\ee{\end{eqnarray}} 
\usepackage[dvipdfmx]{graphicx} 
\usepackage{ulem}
 
\begin{document} 
 
\title{Dielectric response of laser-excited silicon at finite electron temperature} 

\author{S.A. Sato}
\affiliation{Graduate School of Pure and Applied Sciences, University of Tsukuba, Tsukuba 305-8571, Japan}

\author{Y. Shinohara}
\affiliation{Max-Planck-Institut fur Mikrostrukturphysik, Weinberg 2, D-06120 Halle, Germany}

\author{T. Otobe}
\affiliation{Advanced Photon Research Center, JAEA, Kizugawa, Kyoto 619-0615, Japan}

\author{K. Yabana}
\affiliation{Center for Computational Sciences, University of Tsukuba, Tsukuba 305-8577, Japan}
\affiliation{Graduate School of Pure and Applied Sciences, University of Tsukuba, Tsukuba 305-8571, Japan}

\begin{abstract}
We calculate the dielectric response of excited crystalline silicon
in electron thermal equilibrium by adiabatic time-dependent 
density functional theory (TDDFT) 
to model the response to irradiation by high-intensity 
laser pulses. The real part of the dielectric function is
characterized by the strong negative behavior at low frequencies
due to excited electron-hole pairs. The response agrees rather well 
with the numerical pump-probe calculations which simulate 
electronic excitations in nonequilibrium phase immediately 
after the laser pulse irradiation.
The thermal response is also compared with the Drude model 
which includes electron effective mass and collision time
as fitting parameters. 
We find that the extracted effective masses are in the range 
of 0.22-0.36 and lifetimes are in the range of 1-14 fs 
depending on the temperature.  
The short Drude lifetimes show that strong damping is possible 
in the adiabatic TDDFT, despite the absence of explicit 
electron-electron collisions. 
\end{abstract} 
\maketitle 
\section{introduction} 

Properties of dielectrics irradiated by high-intensity and
ultra-short laser pulses have been attracting
substantial interests from both fundamental and technological
points of view \cite{ps99,bk00,cm07,ga11,ba13}.
We are investigating the theory of the dielectric response of materials
to high fields at times shorter than the full local equilibration time.
Time-domain electron dynamics simulation based on the 
time-dependent density functional theory (TDDFT) is 
quite promising for describing the earliest time.
This is the subject of a companion paper, Ref. \cite{paper1}, where
we reported numerical simulations of pump-probe experiments.
This theory should describe the formation of electron-hole excitations
in insulating materials and the energy deposited in them.  At the
next time scale, the electron-hole excitations will come to 
an equilibrium, 
allowing one to treat the system as a thermalized electron-hole 
plasma with fixed numbers of electrons and holes.  
The dielectric properties of this state are the subject of the 
present paper. We will discuss them in crystalline silicon
as in Ref. \cite{paper1}.
At even later times, the atomic degrees of freedom will 
be thermalized as well.  That complete plasma equilibrium is 
beyond the scope of the present work. The two-temperature model
assuming much smaller time-scale of electronic equilibrium than
that of phonons is well established, see Ref. \cite{mu13}.
We will consider the response of thermalized electrons ignoring 
atomic motions, keeping them at equilibrium positions in the 
ground state. This treatment should be reasonable at times 
before substantial part of the electronic excitation energies is 
transferred to lattice motions.

We employ a static density functional theory (DFT) at finite temperature
to describe the thermalized electronic state.
An extension of the DFT to nonzero electronic temperature
was first considered in \cite{me65}, employing the grand canonical
ensemble and introducing a chemical potential for the electrons. 
Recent developments
of finite temperature DFT include discussions on basic aspects of
the theory such as the conditions for the validity  of the adiabatic 
connection formula \cite{pi11} and applications to electrochemical 
reactions \cite{sh08}.
The finite temperature DFT has been applied to the
properties of
matter excited by intense and ultra-short laser pulses.
For example, in Ref. \cite{re06}, lattice properties of laser-excited solids
were investigated using density functional perturbation theory
with the Fermi-Dirac distribution for electrons.
In Ref. \cite{vi10}, finite temperature DFT results were utilized
to analyze solid aluminum excited by XUV pulses.
Our implementation of finite temperature DFT will use the grand 
canonical ensemble for the occupation in the static solution. 
We then calculate the dielectric response in the linear response 
using a real-time method \cite{yb96,be00}. 

Present thermal model calculations are different from previous
numerical pump-probe simulations \cite{paper1} in the population distribution
of electrons. The numerical pump-probe simulations describe
electronic states immediately after the laser irradiation, which
are highly nonequilibrium and anisotropic. 
On the other hand, the present electronic thermal model
describes thermalized, isotropic electronic states.
In spite of these differences, we
will show that many features of response in the elaborated
numerical pump-probe experiments may be reproduced even at a 
quantitative level with the finite temperature calculation,
if we compare two systems at the same number
of excited electrons.
We also compare with a simple Drude response 
embedding the free electrons in a dielectric medium \cite{so00,me10,re10}.  

The construction of the paper is as follows.
In Sec. \ref{sec:thermal}, we describe the theoretical framework of
finite temperature model and present calculated results. 
In Sec. \ref{sec:comparison}, we compare the results with
the Drude model. In Sec. \ref{sec:pump_probe}, we
compare results of the finite temperature model with 
results of numerical pump-probe experiments.
Our findings are summarized in Sec. \ref{sec:summary}.

\section{Electronic thermal model} 
\label{sec:thermal}

\subsection{Ground state}

We model the electronic state of crystalline silicon after 
irradiation of a high-intensity
laser pulse by static DFT for a thermal ensemble of electrons.
Atomic positions are kept at their equilibrium positions in
the ground state, assuming that electron thermalization time
is so short that atomic motions may be ignored.
The Kohn-Sham equation for orbitals is given by
\be
\left\{ -\frac{\hbar^2}{2m_e}\nabla^2 + V_{ion} 
+ \int d\vec r' \frac{e^2}{\vert \vec r - \vec r' \vert} \rho^T(\vec r')
+ \mu_{xc} \right\} \phi_i(\vec r) = \epsilon_i \phi_i(\vec r).
\ee
The electron density at temperature $T$, $\rho^T(\vec{r})$, 
is given by
\be 
\rho^T(\vec{r})=\sum_i n_i^T |\phi_i(\vec{r})|^2 , 
\ee 
where $n_i^T$ is the temperature-dependent occupation number  
of Fermi-Dirac distribution, 
\be 
n_i^T=\frac{1}{1+{\rm e}^{(\epsilon_i-\mu)/k_B T}}.
\ee 
Here $\epsilon_i$ is the energy of electron orbitals, $\mu$ is  
the chemical potential, and $k_BT$ is the temperature in 
energy units. We note that all the quantities related
to the orbitals, $\phi_i$, $\epsilon_i$, and $\mu$ depend
on the temperature $T$ 
due to the self-consistency requirement.

For the present purpose, it is essential to use a functional
which reproduces both indirect and direct band gaps. The reproduction
of the indirect band gap is important to produce correct density of 
electron-hole pairs for a given electronic temperature.
The reproduction of the direct band gap is important for reasonable
descriptions of the optical properties.
We choose the meta-GGA (generalized-gradient approximation) potential 
of Tran and Blaha \cite{tr09} for the exchange-correlation potential,
$\mu_{xc}$. The meta-GGA potential depends on the density $\rho^T(\vec r)$, 
the gradient of the density $|\nabla \rho^T(\vec r)|$, 
and the kinetic energy density $\tau^T(\vec r) = \sum_i n^T_i |\nabla \phi_i(\vec r)|^2$.
The Tran-Blaha meta-GGA potential is known to resolve to some
extent the band gap problem inherent to the local density approximation.
It includes a parameter $c$ to which the band gap is sensitive \cite{ko12}.
We treat it empirically, determining $c=1.04$ which reproduces
the measured indirect band gap of silicon at 1.17 eV.
As will be shown later, the optical gap is also found to be described 
reasonably. The calculated optical gap is about 3.1 eV, in reasonable
agreement with the experimental optical gap, 3.4 eV \cite{zu70}.

Practical calculations are achieved as follows.
We consider only valence electron orbitals treating electron-ion 
interaction by a norm-conserving pseudopotential \cite{tm91,kb82}. 
We use a three-dimensional grid representation to represent
orbital wave functions. The cubic unit cell of a side length
$a=10.26$ a.u. containing eight silicon atoms is discretized
into $20^3$ grid points. The $k$-space is also discretized into
$32^3$ grid points. 

\begin{figure}    
\includegraphics [width = 8cm]{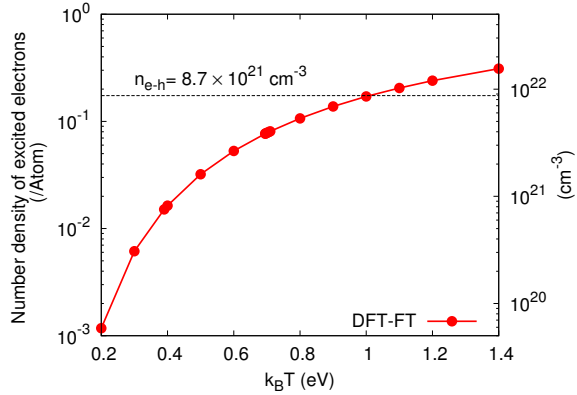}    
\caption{\label{Enex_thermal}
The number density of electron-hole pairs is shown as a 
function of electronic temperature in the thermal DFT
calculation of crystalline silicon. 
} 
\end{figure}

Figure \ref{Enex_thermal} shows number density of 
excited electrons as a function of electron temperature for crystalline
silicon. Here, we define the number density of excited electrons $n_{e-h}$ by,
\be
n_{e-h} = \frac{1}{\Omega}\sum_{i = cond.} n^T_i,
\ee
where the sum is carried out for conduction bands.

As seen from the figure, the number density
of excited electrons monotonically increases as the electron 
temperature increases. 
At electron temperature of 1.0 eV,
which corresponds to 11,600K,
the number density of electron-hole pairs is 0.2 per atom,
indicating excitations of 5 \% of valence electrons.
We note that electronic temperatures and number densities of
excited electrons shown in Fig. \ref{Enex_thermal}
correspond to values of physical interests.
It has been often argued that the critical electron density
is related to the laser damage threshold. The critical electron
density is so defined that the plasma frequency of excited carriers 
coincides with the laser frequency. For Si at $\lambda=625$ nm, 
it is estimated to be $n_c=8.7 \times 10^{21}$ cm$^{-3}$ \cite{so00}.
We also note that several experiments have observed 
laser-excited solids where the number density of excited electrons 
exceeds $10^{22}$cm$^{-3}$ \cite{hu84,so00}.
In theoretical ab-initio calculations, transition of laser-irradiated
silicon into liquid phase has been discussed \cite{si96}.
In the analysis, initial electronic temperature which is necessary
for liquid transition is reported to be 25,000 K (2.15 eV).
In \cite{re06}, instabilities of phonon modes of silicon are reported 
following thermal electronic excitations at temperature 1.5 eV.

\begin{figure}    
\includegraphics [width = 8cm]{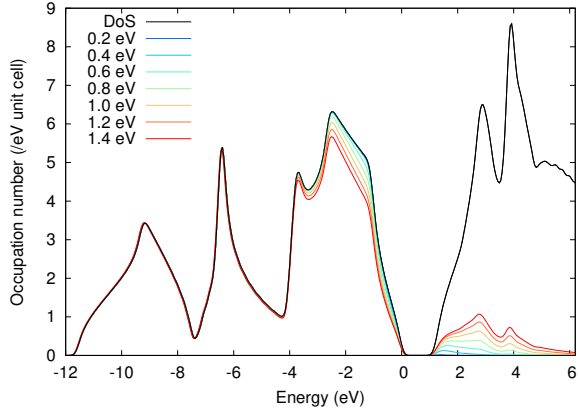}    
\caption{\label{DoS}  
Occupation number distribution of silicon at various temperatures.
The density of states is also shown by black-solid line. } 
\end{figure}

Figure \ref{DoS} shows occupation distributions at various 
temperatures, as well as the density of states shown by
black solid line. At temperatures around 1 eV, we find a 
substantial excitations of electrons from orbitals within 3 eV 
below the highest occupied orbital to orbitals within 5 eV 
above the lowest unoccupied orbitals.
From the figure, we find that there is little change in the
amount of band gap for wide temperatures. 
In literatures \cite{be01,fa06}, changes of band gap due
to band gap renormalization effect \cite{fa06} and
to a decrease of electron-hole attraction \cite{be01}
have been investigated. They are originated from
screening effects by excited carriers. We consider that
these correlation effects are not properly treated
in our thermal TDDFT calculation with meta-GGA potential.

\subsection{Linear response}

We  calculate dielectric properties of the medium
in the adiabatic TDDFT, using the same
Tran-Blaha meta-GGA potential for the response calculation.
Numerically, we solve the following time-dependent Kohn-Sham
equation in real time to calculate the dielectric property, 
\be
i\hbar \frac{\partial}{\partial t} \psi_i(\vec r,t)=
\left\{ \frac{1}{2m_e}\left( \vec p + \frac{e}{c} \vec A(t) \right)^2 + V_{ion} 
+ \int d\vec r' \frac{e^2}{\vert \vec r - \vec r' \vert} \rho(\vec r',t)
+ \mu_{xc}  \right\} \psi_i(\vec r,t).
\label{eq:tdks}
\ee
The density $\rho(\vec r,t)$ is constructed from time-dependent
orbitals as $\rho(\vec r,t) = \sum_i n^T_i |\psi_i(\vec r,t)|^2$, 
using the occupation numbers in the ground states. 
To explore the dielectric property,
we apply a distorting vector potential of step function in time \cite{be00,ya12}. 
\be
\vec A(t) = \vec e_{\beta} A_0 \theta(t),
\ee
where $\vec e_{\beta}$ is a unit vector in the $\beta$ direction.
We calculate the current flowing within the unit cell from the
solution by
\be
\vec J(t) = -\frac{e}{\Omega} \sum_i n^T_i \int_{\Omega} d\vec r d\vec r'
\psi_i^*(\vec r,t) \vec v(\vec r,\vec r') \psi_i(\vec r',t),
\ee
where $\Omega$ is a volume of the unit cell and the velocity
operator $\vec v(\vec r,\vec r')$ is defined by
\be
\vec v(\vec r,\vec r') = -\frac{i\hbar}{m_e} \vec\nabla
\delta(\vec r,\vec r') + \frac{1}{i\hbar}
\left[ \vec r V_{ps}^{NL}(\vec r,\vec r')-V_{ps}^{NL}(\vec r,\vec r')\vec r'
\right],
\ee
where $V_{ps}^{NL}$ is the nonlocal part of the pseudopotential.
The conductivity is calculated from the induced current by
\be
\sigma_{\alpha\beta}(\omega) = -\frac{c}{A_0}
\int_0^T dt e^{i\omega t} W(t/T) J_{\alpha}(t),
\ee
where $J_{\alpha}(t)$ is the $\alpha$ component of $\vec J(t)$,
and $T$ is the duration of time evolution. 
We use the mask
function $W(x)$ given by $W(x)=1-3x^2+2x^3$ \cite{ya06}.
The dielectric function is obtained from the conductivity by
\be
\epsilon_{\alpha\beta}(\omega)
=\delta_{\alpha\beta}+
\frac{4\pi i \sigma_{\alpha\beta}(\omega)}{\omega}.
\ee
In silicon, only diagonal element appears in the thermal model, 
$\epsilon_{\alpha\beta}(\omega) = \delta_{\alpha\beta}\epsilon(\omega)$.

In time evolution calculations, we use the same grid points in
the real space and the $k$-space as those in the static calculation. 
The time propagation is computed 
using a fourth-order Taylor expansion method \cite{yb96}, 
with a time step of $\Delta t=$ 0.04 a.u.
The total duration of the time evolution is $T=1,280$ a.u. with
the number of time steps $N_T=32,000$. 

\begin{figure} 
\includegraphics [width = 8cm]{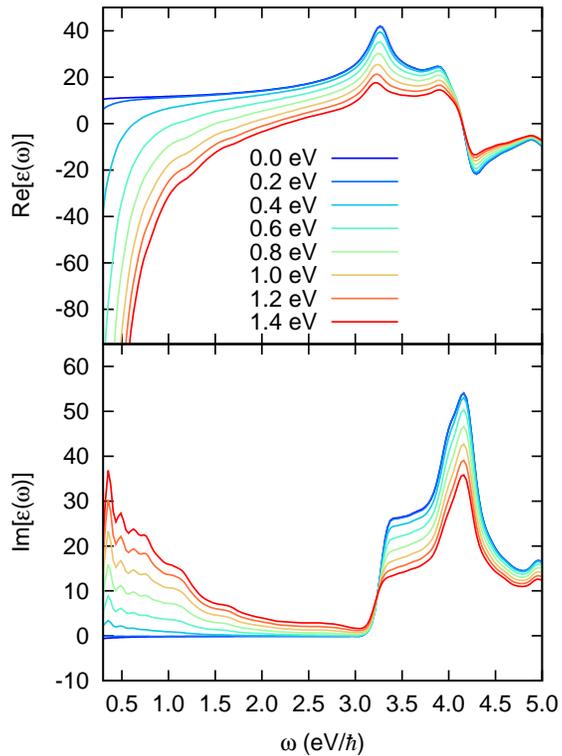}    
\caption{\label{eps_thermal} 
The dielectric function of silicon in the finite temperature model  
at several temperatures. Top panel shows the real part of  
the dielectric function, and the bottom panel shows the  
imaginary part.  
} 
\end{figure}

In Fig. \ref{eps_thermal}, we show dielectric functions of  
silicon at several electron temperatures.  In the real part of 
the dielectric function,  all responses at finite temperatures 
show a strong negative behavior at low frequencies.
This Drude-like behavior comes from excited electron-hole pairs.
The low energy component of the imaginary part shows absorptive 
contributions at low frequencies, increasing monotonically 
as the temperature increases. 
In our previous study employing numerical pump-probe
experiments \cite{paper1} which catch nonequilibrium
distributions of electron-hole pairs, we have observed 
a similar behavior of Drude-like divergence in the real part. 
However, the absorptive contribution in the imaginary part
was not observed.
 
\begin{figure} 
\begin{center}
\includegraphics [width = 8cm]{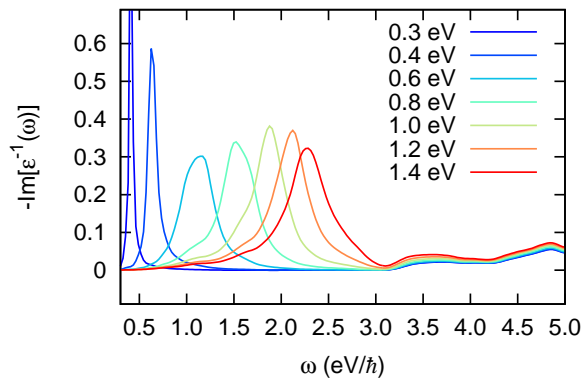}    
\caption{\label{ineps_thermal}  
The imaginary part of the inverse dielectric function for 
various electronic temperatures. 
} 
\end{center}
\end{figure}

A convenient way to exhibit the plasmon contribution to the response
is to plot the imaginary part of the inverse dielectric function, ${\rm Im} 
\epsilon^{-1}$.  This is shown in Fig. \ref{ineps_thermal} for
several temperatures up to $k_B T = 1.4$ eV.  At the lowest
temperature, one sees a very sharp plasmon peak, located at
an energy of $\sim 0.4$ eV.  The plasmon excitation energy 
increases with temperature, due to the increased density of 
electron-hole pairs. 
We note that the width of the plasmon also increases with  
temperature, up to about $k_B T \approx 0.6 $ eV.  Beyond that,
the width does not change very much, up to the maximum temperature
considered.

We note that local field corrections are not important in
the above results. Namely, results showns above hardly change
if we fix the Kohn-Sham Hamiltonian in Eq. (\ref{eq:tdks}) to that in 
the thermal ground state.
 
\section{Comparison with free-carrier models} 
\label{sec:comparison}
 
The dielectric response of solids excited by intense and  
ultrashort laser pulses is often modeled by a simplified  
dielectric function, adding a Drude-like component to the  
dielectric function in the ground state \cite{me10,re10}. In this section,  
we will show that a model of this kind can reproduce quite
well our calculated finite-temperature response.
 
We consider a model proposed by Sokolowski-Tinten and von der Lind  
\cite{so00}, which we shall call the SL model. 
They consider three physical effects for the dielectric response
of laser-excited semiconductor: (i) state and band filling,
(ii) renormalization of the band structure, and (iii) the
free-carrier response.
The SL dielectric function is parameterized as 

\be 
\epsilon_{SL}(\omega) &=& 1+\left [ 
\epsilon_0 (\omega+\Delta E_{gap})-1 \right ] 
\frac{n_0-n_{eh}}{n_0} \nonumber \\ 
&&-4\pi \frac{e^2n_{eh}}{m^*\omega(\omega+i/\tau)}.
\ee 
Here $\epsilon_0(\omega)$ is the dielectric function in the ground
state for which we employ the one calculated at 
zero temperature. $\Delta E_{gap}$ is the change of the band gap 
by the laser irradiation for which the calculated shift of the 
gap energy is used. $n_{eh}$ is the electron-hole pair density
for which we use the calculated values. Three other parameters are: 
the effective mass $m^*$; the Drude damping time $\tau$; and the active
number of valence electrons $n_0$. These are treated as fitting parameters. 
  
\begin{figure} 
\includegraphics [width = 7.5cm]{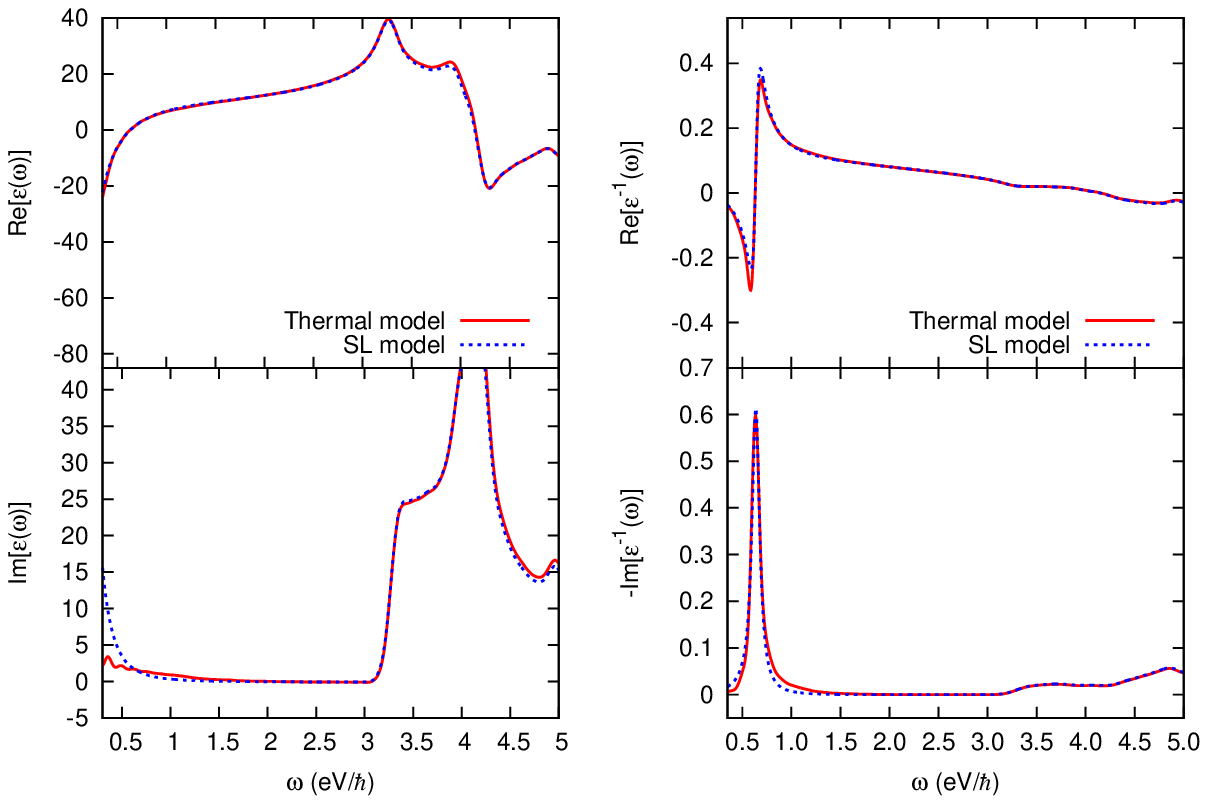}
\includegraphics [width = 7.5cm]{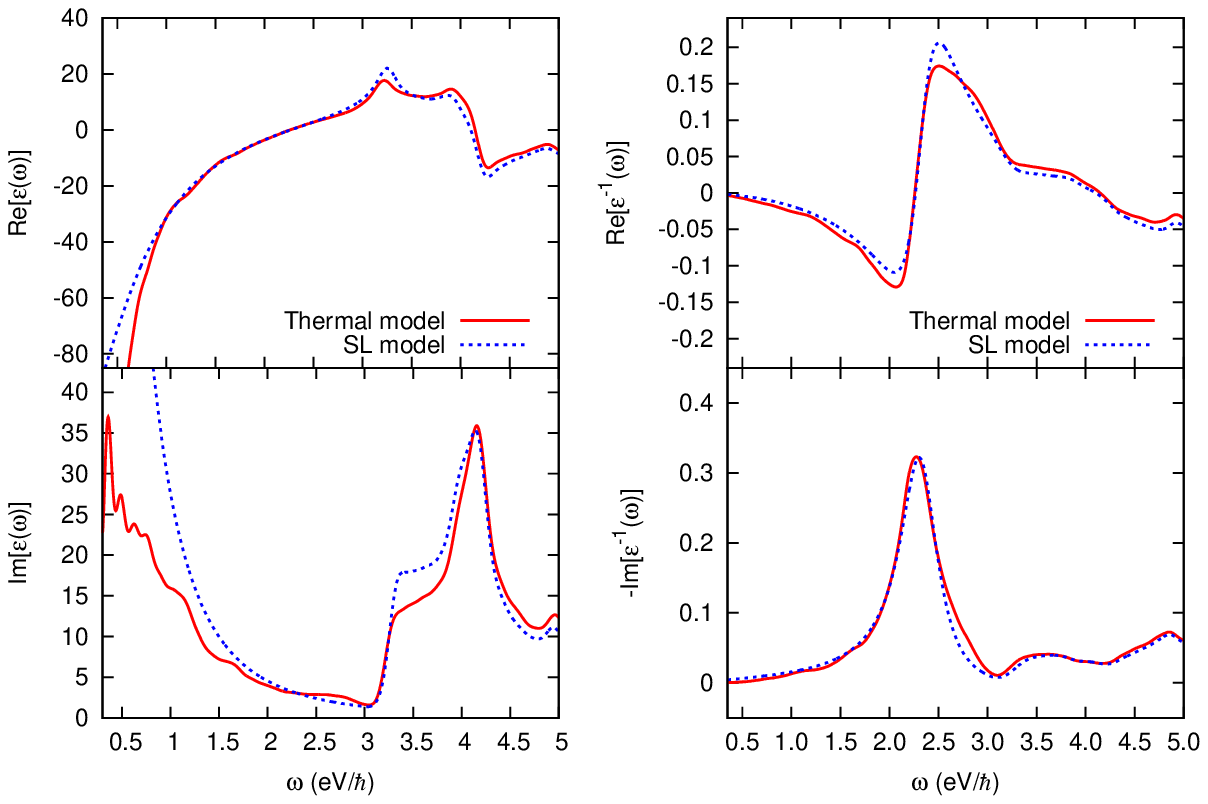}    
\caption{\label{eps_thermal_drude}  
Comparison of the thermal model and a fit with the 
SL model. The electronic temperature in the thermal 
model is $k_B T = 0.4$ eV (left) and 1.4 eV (right). 
} 
\end{figure}

The fit is carried out by minimizing the mean square
error as given by
\be 
I_{error}=\int_{\omega_i}^{\omega_f}d\omega \left| 
\epsilon_{T}^{-1}(\omega)-\epsilon_{SL}^{-1}(\omega) 
\right|^2, 
\ee 
where $\epsilon_{T}(\omega)$ is the dielectric function in the
thermal model. We take the interval $\hbar \omega_i= 0.3$ eV 
and $\hbar \omega_f=6.0$ eV. The quality of the fit is
shown in Fig. \ref{eps_thermal_drude}  for temperatures
of $k_B T= 1.4$ and 0.4 eV in the thermal model.  The fit is very good
except for the ${\rm Im} \epsilon$ at the lowest
frequencies. In particular, the  
plasmon peak in the inverse dielectric function is very
well reproduced.
 
\begin{figure} 
\includegraphics [width = 8cm]{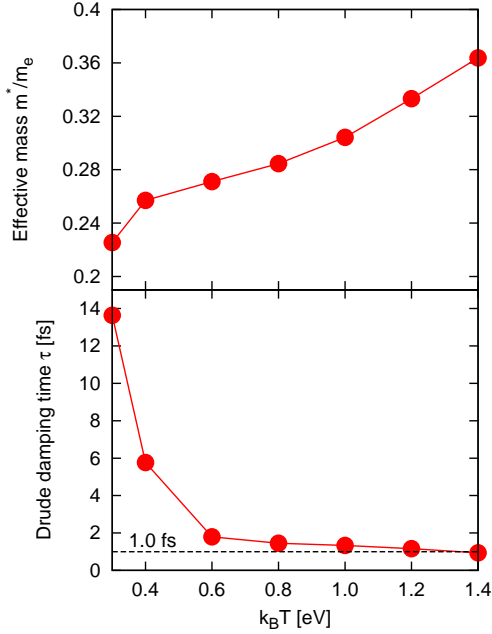} 
\caption{\label{mtau_thermal} 
Parameters of the SL model determined by the fitting procedure 
to the thermal model. 
Top panel shows the effective mass $m^*$ and the bottom panel  
shows the Drude damping time $\tau$. } 
\end{figure}

In Fig. \ref{mtau_thermal}, we show the fitted effective mass 
$m^*$ and the collision time $\tau$ as functions of the temperature 
in the thermal model. 
The top panel shows that the effective mass $m^*$ increases
as the temperature increases. We have found a similar behavior
in the numerical pump-probe experiments in Ref. \cite{paper1}.
The change of effective mass may be understood by the 
change of the distribution of the electron-hole pairs
in $k$-space. 
 
The bottom panel of Fig. \ref{mtau_thermal} shows that the damping 
time $\tau$ becomes very small as the electron temperature increases. 
The value of $\tau$ monotonically decreases and reaches a value of 
1.0 fs at $k_B T \approx 1.4$ eV.  At first sight this is puzzling, 
because there are no explicit collision effects in either the TDKS equation 
or in the thermal model in the adiabatic meta-GGA which we adopted.
Since we fix ion positions during time evolution
calculations of orbitals, no electron-phonon interactions are taken
into account.
In spite of them, our plasmon peak has a large damping, 
corresponding to collision times as short as 1.0 fs in the thermal model. 
We consider that the damping arises from the elastic scattering of electrons 
from ionic core  potentials. Since the electron-ion interactions
constitute periodic potential for electrons, we may equivalently
say that the damping is due to the interband transitions of
excited carriers.
We note that TDDFT treatment of linear response  describes
the dielectric function of metals fairly well, including the width
of plasmon seen in the inverse dielectric function \cite{be00}.

\section{Comparison with numerical pump-probe experiments} 
\label{sec:pump_probe}

In the preceding paper \cite{paper1}, we have carried out numerical 
pump-probe experiments to extract dielectric properties of laser-excited
silicon immediately after irradiation by the laser pulse. 
This method catches fully the nonequilibrium nature of the excited
electrons. The difference between the numerical pump-probe calculations  
and the present thermal model comes entirely from the different 
electron-hole distributions in the excited system to be probed. 
In this subsection, we compare their predicted dielectric functions.

In the numerical pump-probe calculation, we solve the TDKS equation 
in real time where the electric fields of both pump and probe pulses 
are included. The pump electric field $E_P(t)$ excites electrons and 
probe electric field $E_p(t)$ is used to extract dielectric properties 
of excited silicon. The dielectric properties are examined from
the currents induced by the electric fields. 
In practice, we performed two calculations. In one calculation,
we include both pump and probe electric fields, $E_P(t)+E_p(t)$,
in the TDKS equation. We denote the current in 
this numerical pump-probe calculation as $J_{Pp}(t)$.
The other calculation includes only the pump field $E_P(t)$ and we 
denote the current as $J_P(t)$. The difference of the currents,
$J_p(t)=J_{Pp}(t)-J_P(t)$ brings information of excited silicon.
The electric conductivity $\sigma(\omega)$ of excited silicon is given by
\be
\sigma(\omega) = \frac{\int dt J_p(t) e^{i\omega t}}{\int dt E_p(t) e^{i\omega t}},
\ee
and the dielectric function by $\epsilon(\omega)=1+4\pi i \sigma(\omega)/\omega$. 
In the numerical pump-probe experiment,
we note that the responses are not isotropic but depend on the angle
between electric fields of pump and probe fields. We consider
two cases: the pump and probe electric fields are parallel
and perpendicular to each other.

To compare results of the thermal model with those of
the numerical pump-probe experiments, we first need 
to assume a correspondence between the excited systems 
that we wish to compare.  
Since the plasmon characteristics are closely tied to the number of 
electron-hole pairs, we shall use that measure to make the comparison.  

\begin{figure}     
\includegraphics [width = 8cm]{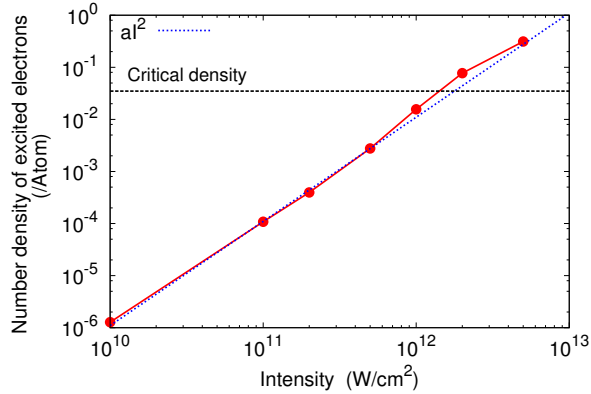}     
\caption{\label{Enex_I}   
The number density of electron-hole pairs of the crystalline 
silicon in the final state following the pulsed excitation 
as a function of the maximum pump intensity determined as $I=cE_0^2/8\pi$. 
The critical density is indicated by the horizontal line. 
The squared intensity line normalized at 10$^{10}$ W/cm$^2$ is 
also shown by blue-dotted line. Taken from \cite{paper1}.} 
\end{figure}    

In Ref. \cite{paper1}, we reported calculations solving 
the TDKS equation with the electric field of the applied laser 
pulse whose vector potential is given by
\begin{eqnarray}  
\label{pump-laser}  
A(t) &=& \nonumber \left\{ \begin{array}{ll}  
    -c\frac{E_0}{\omega_P} \cos{(\omega t)}\sin^2(\pi t/{\tau_L})   
& (0<t<{\tau_L}) \\  
    0 & ({\rm otherwise}) ,
  \end{array} \right. \nonumber \\  
\end{eqnarray} 
where $\omega$ and $\tau_L$ is the average frequency and 
the time length of the laser pulse, respectively.
$E_0$ is the maximum electric field strength in the medium.
We denote the maximum intensity of the pulse given by
$I=cE_0^2/8\pi$. 
Using the laser pulse of the frequency $\hbar\omega=1.55$ eV and
the duration of the pulse $\tau_L=18$ fs, the number density
of excited electrons is calculated for laser pulses of
several intensities. We show the result in Fig. \ref{Enex_I} 
which is taken from \cite{paper1}.
Combining Fig. \ref{Enex_I} and Fig. \ref{Enex_thermal},
we can relate the laser intensity $I$ and the electronic temperature
$k_B T$ through the number density of electron-hole pairs $n_{eh}$.
For example, in the TDKS calculation using the laser pulse of
$I=1.0 \times 10^{12}$W/cm$^2$ excites electron hole pairs of
$n_{eh}=0.016 /$Atom.  From Fig. \ref{Enex_thermal}, the
corresponding temperature is given by $k_B T=0.4$ eV.
For the laser pulse of $I=5.0 \times 10^{12}$W/cm$^2$, 
the density of electron-hole pair is $n_{eh}=0.31 /$Atom.
Corresponding temperature is $k_B T=1.4$ eV.
In the following, we use $n_{eh}$ to specify calculations of 
the finite temperature model and the numerical pump-probe 
experiments.

\begin{figure}
\includegraphics [width = 10cm]{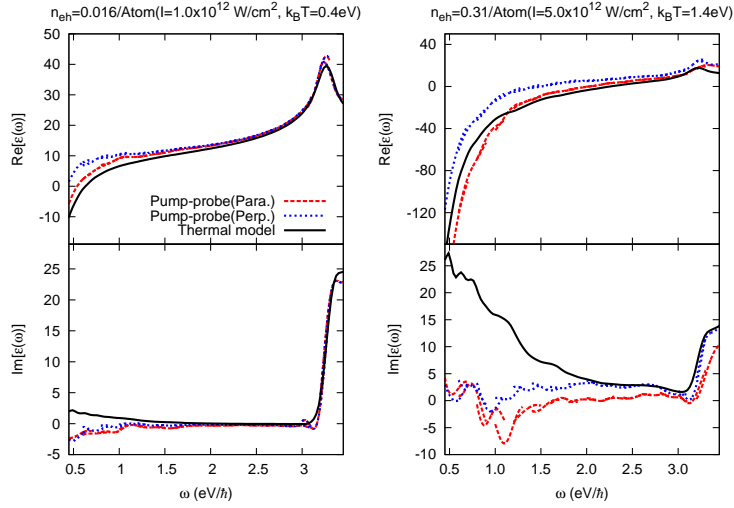}    
\caption{\label{eps_pp_thermal}  
Comparisons of the dielectric function of the numerical pump-probe 
calculation \cite{paper1} and the thermal model.  Left-hand panels:  
$n_{eh} = 0.016$ /Atom; right-hand panels:  $n_{eh} = 0.31$ /Atom.
} 
\end{figure}

We show a comparison of dielectric function by two methods for two cases,
$n_{eh}=0.016$ /Atom and $n_{eh}=0.31$ /Atom, in Fig. \ref{eps_pp_thermal}. 
The black lines show dielectric function of thermal model.
The red-dashed line and the blue-dotted line show the
results of the numerical pump-probe calculations for probe polarization
parallel and perpendicular to the pump, respectively.

As seen from the figure, the real part of the dielectric function of 
silicon excited by the pump pulse is close to the thermal model
for two cases. At lower excitation of $n_{eh}=0.016$ /Atom,
the thermal model is close to the pulsed excitation in the
parallel probing. At higher excitation of $n_{eh}=0.31$ /Atom,
the thermal model is again close to the pulsed excitation in
the parallel probing at higher frequencies ($\hbar\omega > 1$ eV)
and is between the parallel and perpendicular probings at
low frequencies ($\hbar\omega < 1$ eV).
The imaginary part of the dielectric function looks rather different. 
While the thermal model predicts positive imaginary part below 
the band gap, the pulse-excited silicon shows much smaller value, 
even negative in certain frequencies.

The difference between two calculations comes entirely from different
distributions of electron-hole pairs: thermal equilibrium
distributions in the thermal model and nonequilibrium distributions
in the numerical pump-probe simulation. 
To clarify the difference, we investigate population 
distributions in energy and momentum space.

We first denote the orbital index $\{ i \}$ in terms of
$\{ b, \vec k \}$, where $b$ indicates bands and $\vec k$ indicates
the Bloch momentum. Occupation numbers are expressed as
$n_{b\vec k}^{X}$, where $X=T$ for thermal model and $X=NPP$ for
numerical pump-probe simulation.
We define the occupation distribution function by
\be
f^X(\vec k, \epsilon) = 
\sum_{b} n_{b \vec k}^X \delta(\epsilon - \epsilon_{b \vec k}^X).
\ee
For numerical pump-probe simulation, we define the energy
eigenvalue $\epsilon^{NPP}_{b\vec k}$ by solving the following
Kohn-Sham equation,
\be
\hat h_{KS}^{NPP}(t_f) \phi_{b\vec k}^{NPP} = \epsilon_{b\vec k}^{NPP} \phi_{b\vec k}^{NPP},
\ee
where $h^{NPP}_{KS}(t_f)$ is the time-dependent Kohn-Sham Hamiltonian
at time $t_f$ when the laser pulse ended.
The occupation number in the numerical pump-probe simulation is defined by
\be
n_{b\vec k}^{NPP} = \sum_{b' \vec k'}
\vert \langle \phi_{b\vec k}^{NPP} \vert \psi_{b'\vec k'}^{NPP}(t_f) \rangle \vert^2,
\ee
where $\psi_{b\vec k}^{NPP}$ is the solution of Eq. (\ref{eq:tdks}) 
at time $t_f$.

\begin{figure}
\includegraphics [width = 8cm]{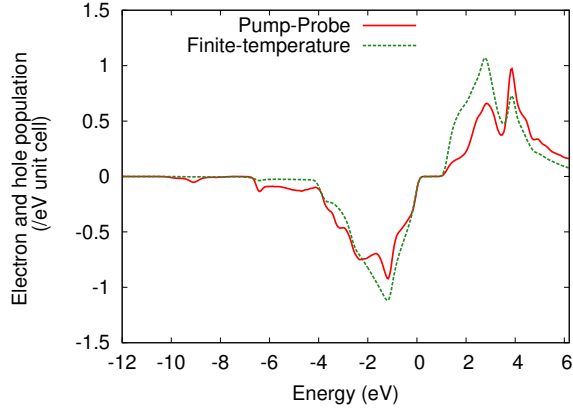}    
\caption{\label{Si_DoS_pp_ft}
Population distribution of electrons and holes 
in laser-excited silicon.
} 
\end{figure}

Using the occupation distribution function, we first calculate the occupation distribution 
as a function of energy,
\be
D^X(\epsilon) = \sum_{\vec k} \left\{ f^X(\vec k,\epsilon) -f^0(\vec k,\epsilon) \right\},
\ee
where $f^0(\vec k,\epsilon)$ is the occupation distribution function
in the ground state at zero temperature.
The calculated results are shown in Fig.  \ref{Si_DoS_pp_ft} for cases when 
$n_{e-h}=0.31$ /Atom. Red-solid line shows the distribution of 
the numerical pump-probe simulation, and green-dotted line shows
that of the thermal model. We set the highest energy of the valence band to zero. 
Positive values at positive energy region show distribution of electrons in
conduction band, while negative values at negative energy region show
the hole distribution in the valence band.

From the figure, we observe that electrons and holes distribute
in wider energy region in the numerical pump-probe simulation than 
those in the electron thermal model.
The decrease of the lower energy electron-hole and the increase of 
higher energy electron-hole in the numerical pump-probe simulation
may cause optical emissions which negatively contribute to the
imaginary part of the dielectric function. This explains small or even 
negative values of the imaginary part of the dielectric function
in the numerical pump-probe simulation.

To further clarify the difference in electron-hole distributions,
we calculate the distribution in the Bloch momentum space.
We note that the Bloch momentum does not correspond to that in 
the primitive cell since we employ the cubic unit cell containing
eight silicon atoms in our calculation.
We define the distribution of electrons in the following way:
\be
D^X_{e}(\vec k) = \int^{\infty}_0 d \epsilon 
\left\{ f^X(\vec k,\epsilon) - f^0(\vec k,\epsilon) \right\}.
\ee
For the distribution of holes, integration is achieved for 
$\epsilon < 0$.
We note that there holds $D^{NPP}_e(\vec k) = -D^{NPP}_h(\vec k)$.

\begin{figure}
\includegraphics [width = 9.5 cm]{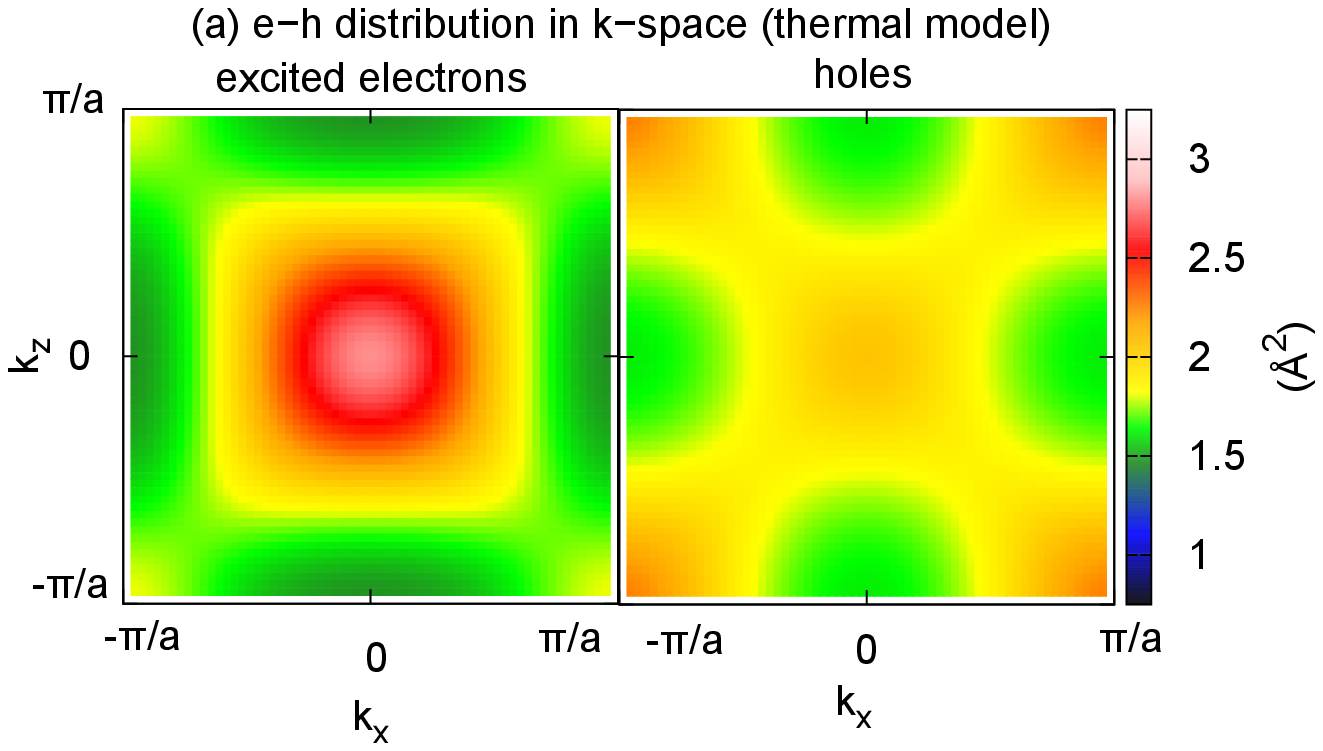}    
\includegraphics [width = 6cm]{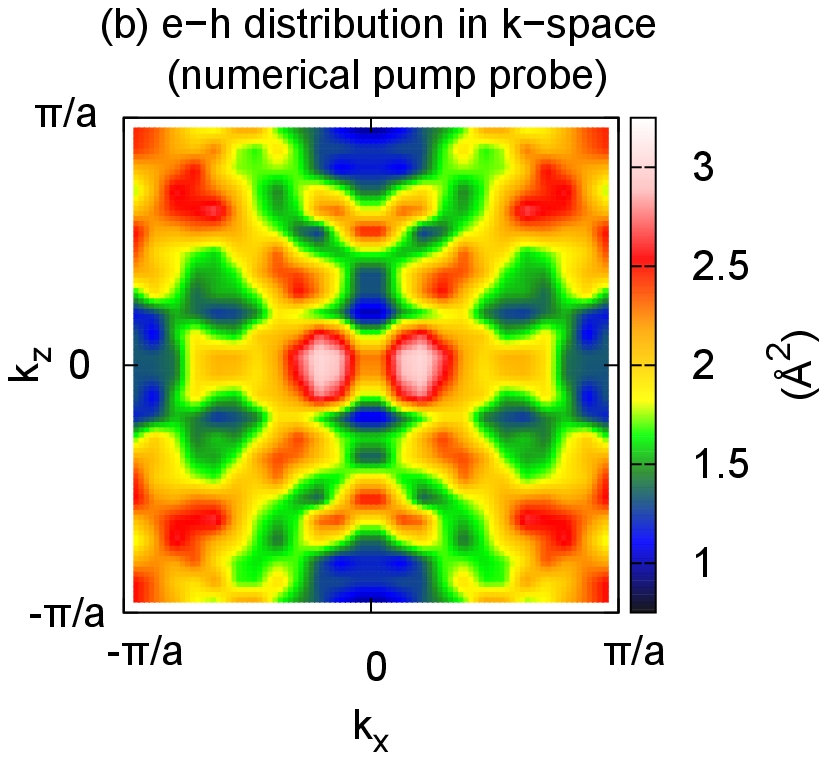}    
\caption{\label{nex_dist_k_ft_mod}    
Population distributions of excited electrons and holes 
in the Bloch momentum space. The panel (a) shows 
the excited electron (left hand side) and hole (right hand side)
distributions in the thermal model at electron temperature
$k_BT=1.4$eV, while the panel (b) shows the excited electron
distribution in the numerical pump-probe
method at the pump intensity $I=5.0\times10^{12}$W/cm$^2$.
The population distributions are shown in the $k_x$-$k_z$ plane
integrating over $k_y$.
} 
\end{figure}

Figure \ref{nex_dist_k_ft_mod} (a) shows the distributions of electrons
and holes in the thermal model at electron temperature $T=1.4$eV, while 
Figure \ref{nex_dist_k_ft_mod} (b) shows the distribution of
electrons in the numerical pump-probe simulation at the pump intensity 
$I=5.0\times10^{12}$W/cm$^2$. The polarization direction
of the pump pulse is set parallel to $z$-direction. 
In both panels, distributions integrated over $k_y$ are shown 
in $k_x$-$k_z$ plane.
As is evident from panels (a) and (b), there is a large difference 
in the distribution in momentum space between the thermal model 
and the numerical pump-probe simulation. In the thermal model, distributions
of electrons and holes are different, reflecting the indirect band gap structure. 
The distribution in the numerical pump-probe 
simulation shows much more complex, structured, and nonuniform behavior 
than that in the thermal model,
since electrons and holes are in nonequilibrium phase immediately 
after the laser irradiation. 
We note that the real parts of the dielectric
functions do not show large differences between two calculations
(See Fig. \ref{eps_pp_thermal}). This indicates that the real part of 
dielectric function is sensitive to the number density of excited electrons, 
not to the detailed distribution of electrons and holes.

We thus conclude that the thermal model describes the real part
of the dielectric function quite well, provided the number density of
electron-hole pairs is the same. The difference 
between two calculations comes from the nonthermal distribution
of electron-hole pairs in numerical pump-probe simulation. It seems that the difference 
is more evident for the imaginary part. A nonequilibrium phase
of electronic excitations manifests more sensitively in the
imaginary part of the dielectric function.

\section{SUMMARY} 
\label{sec:summary}

We investigated the change of dielectric response  induced
by intense and ultrashort laser pulses by a thermal model,
assuming electronic equilibrium.
This description is expected to apply to excited matter 
after a few tens of femtosecond following the laser irradiation.
We first solved the static Kohn-Sham equation with finite
temperature Fermi-Dirac function occupation factors.  
Its dielectric response was then computed by applying 
the linear response theory using the real-time method.  

The calculated thermal dielectric function is characterized by the 
strong negative behavior in the real part at low frequencies 
caused by excited electron-hole pairs. The imaginary part
shows absorptive contributions at low frequencies,
increasing monotonically as the temperature increases.
Plotting it in the inverse dielectric function, a sharp plasmon
feature manifests clearly. The frequency of the plasmon
increases monotonically with temperature due to the increased
density of electron-hole pairs. The width also increases 
at low temperature region, then becomes almost constant.

The thermal dielectric function is compared with a simple
Drude model of free-electron dynamics, embedded in the
dielectric medium corresponding to the ground state.
There are three basic parameters determining the 
electron-hole plasma properties, namely the density of 
electron-hole pairs, their effective mass $m^*$, and the 
collision time $\tau$.
The density of electron-hole pairs is known from the thermal
ground state calculation, but the other quantities are fit.
We find the collision time of as short as 1.0 fs gives  
reasonable fit. This short value for the collision time  
is unexpected, since there are no explicit collision terms
in the time-dependent Kohn-Sham equation that we solve.
We consider the short collision time comes from the  
elastic scattering of electrons from atoms.
 
We also compared the thermal dielectric function with
that derived from numerical pump-probe calculation in which
electronic response is derived from time evolution of
Kohn-Sham orbitals under electric fields of both pump and probe pulses.
The numerical pump-probe simulation describes the response
of excited matter in the nonequilibrium state reached just after
the pulse has been applied.  We find the real part of the
dielectric function shows reasonable correspondence
if we compare them at the same number density of electron-hole
pairs. However, the imaginary part shows marked difference.
The thermal dielectric function shows positive imaginary
part, while the numerical pump-probe calculation gives small
contribution in the imaginary part, even negative contribution.

The above difference comes from the distributions of electrons and 
holes.
To clarify the origin of the difference, we investigated
the distribution in energy and momentum space. 
From the population distribution in energy domain, 
we found that electrons and holes distribute in wider energy region 
in the numerical pump-probe simulation than those in the thermal model.
The decrease of the lower energy electron-hole and the increase of 
higher energy electron-hole in the numerical pump-probe simulation
may cause optical emission which negatively contribute to the
imaginary part of the dielectric function. This explains small or even 
negative values of the imaginary part of the dielectric function
in the numerical pump-probe simulation.
From the population distribution in Bloch momentum, we found 
large differences between the thermal model and the pump-probe simulation.
The distribution in the numerical pump-probe simulation is much structured and 
nonuniform compared with the thermal case, reflecting nonequilibrium
phase immediately after the end of the incident pulse. 

In spite of the large difference of the electron-hole distributions 
between the thermal model and the numerical pump-probe simulation, 
the real parts of the dielectric functions are qualitatively similar.
Moreover, the real parts of the dielectric functions in both cases can be 
well described by the Drude model. 
The real part of the dielectric is well described by the Drude model
using only the number density of excited-electrons and the effective mass. 
Therefore, we may validate the estimation of the number density of 
excited-electrons in laser-excited solids using the Drude model for 
both non-equilibrium and thermal phases, based on the microscopic 
treatment of the quantum electron dynamics.

\section*{Acknowledgments} 
 
We thank G.F. Bertsch for discussions and suggestions.
This work is supported by the Grants-in-Aid for Scientific Research No.
23340113, No. 23104503, No. 21340073, and No. 21740303. The numerical
calculations were performed on the supercomputer at the Institute of Solid State Physics, University of Tokyo, and T2K-Tsukuba at the 
Center for Computational Sciences, University of Tsukuba.

\end{document}